\documentclass[twocolumn,aps,showpacs,prc]{revtex4}
\usepackage[dvips]{epsfig}

\begin{document}

\title{Direct photons from Au+Au collisions at RHIC: QGP vs. hot
 hadronic gas}

\author{A. K. Chaudhuri}
\email[E-mail:]{akc@veccal.ernet.in}
\affiliation{Variable Energy Cyclotron Centre,\\ 1/AF, Bidhan Nagar,
Kolkata 700~064, India}

\begin{abstract}
We have analysed the preliminary PHENIX data on the transverse momentum distribution of direct photons in 0-20\% centrality Au+Au
collisions at $\sqrt{s_{NN}}$=200 GeV.
In ideal hydrodynamics, data are explained if Au+Au collision produces
Quark-Gluon-Plasma at the temperature $T_i$=400 MeV, at an initial time $\tau_i$=0.6 fm. PHENIX data are not explained  in the alternate
scenario when Au+Au collisions produces  hot hadronic gas at temperature
$T_i \leq$  220 MeV at an initial time $\tau_i  \leq$  5 fm. 
\end{abstract} 


\maketitle

In Au+Au collisions at RHIC, one observe a dramatic suppression 
of high $p_T$ hadrons \cite{BRAHMSwhitepaper,PHOBOSwhitepaper,PHENIXwhitepaper,STARwhitepaper} . The suppression is more in central  than in peripheral collisions.
It is also established that high $p_T$ suppression is a final state effect,
a parton before fragmenting in to hadrons suffers energy loss in a
dense matter, leading to suppressed production
\cite{Gyulassy:2003mc}. High $p_T$ suppression together with the 
observation that bulk of the hadron production data ($p_T$ spectrum 
and elliptic flow etc.) are well described in a {\em ideal} hydrodynamic 
model \cite{Kolb:2003dz,Kolb:2001qz}, strongly support the idea that 
RHIC has produced  thermalised matter at very high energy density. 
However it is not certain whether the matter produced is strongly interacting 
Quark-Gluon-Plasma (sQGP) as predicted in lattice QCD calculations.  
Hadrons, being strongly interacting, are emitted from the  surface of the thermalised matter and
carry information about the freeze-out surface only. They are unaware of
the condition of the interior of the matter and can provide information about the deep interior only in an indirect way.
In a hydrodynamic model, one fixes the initial conditions of the fluid  such that the "experimental" freeze-out surface is correctly reproduced.  
In contrast 
to hadrons,    photons being weakly interacting, are emitted from 
whole volume of the matter. Throughout the evolution of the matter, photons are emitted. Conditions of the produced matter, at its deep interior,
are better probed by the photons.
 
Recently  PHENIX collaboration published their measurements of direct photons in different centrality ranges of Au+Au collisions at $\sqrt{s_{NN}}$=200 GeV \cite{Adler:2005ig}.  pQCD model calculation of 
direct photons in p+p collisions, multiplied with the thickness function, correctly reproduces the data in all the centrality ranges of collisions. Transverse momentum
distribution of direct photons in different centrality ranges of collisions
scale with the number of binary collisions. Apparently the direct photon results are in direct contradiction with high $p_T$ suppression observed 
in Au+Au collisions.   {\em Partons fragmenting in to photons do not suffer 
energy loss in contrast to partons  fragmenting in to hadrons}. Moreover, 
very good description of the published data with pQCD photons donot leave any room for thermal photons, which are expected to be emitted in 
large numbers from the thermalised matter.

Measurement of direct photons is very challenging, more so at low 
$p_T$. The huge background from $\pi^0$ decay, dominate the 
spectra and proper algorithm for background subtraction is very 
important.  Very recently PHENIX collaboration improved upon the
analysis technique for photon measurement. 
Compared 
to the  conventional method
the new method improves both the signal to background ratio and the energy resolution. In QM2005, PHENIX collaboration presented 
the (preliminary) result of their new analysis \cite{Buesching:2005pb}.
With  the new analysis technique, direct photon yield in the low $p_T$ range 
(1-4 GeV), in 0-20\% centrality Au+Au collisions is increased substantially.
pQCD predictions no longer can explain the data. Direct photons in excess of pQCD (hard) photons are possibly   from a  thermal source.  The PHENIX (preliminary) direct photon data thus provides the first opportunity to measure the initial condition of the matter   produced in RHIC Au+Au collisions, at deep interior.

In the present paper, in a hydrodynamic model,    we have analysed 
the PHENIX (preliminary) direct photon data.  Procedure for obtaining photon spectra in a hydrodynamic evolution is well known \cite{VonGersdorff:1986yf}. We have solved the hydrodynamic 
equations $\partial _{\mu}T^{\mu \nu }=0$ for a baryon free gas assuming cylindrical symmetry and boost-invariance. We have considered two possible scenarios: (i) Au+Au collisions produce QGP as the initial
state and (ii) Au+Au collisions produces hot hadronic gas. In the 
first scenario produced QGP expands,  cools,  undergoes 1st order phase transition at critical temperature ($T_c$), enters a mixed phase, remain  in  the  mixed
phase  till  all  the  quark  matter is converted into a hadronic matter then cools to freeze-out temperature. In the second scenario the
hot hadronic gas expands, cools till the freeze-out. In both the scenarios,
photons are emitted throughout the evolution and their yield in integrated over the space-time volume.
The second   scenario is very important
for unambiguous detection of QGP. Direct photons are very strange probe.
Theoretical predictions \cite{Turbide:2003si} indicate that they are emitted {\em equally well} from the QGP and from the hot hadronic phase. Only in a narrow transverse momentum window around 3 GeV, the two phases may
be distinguished. Indeed, direct photons measured at SPS energy in S+Au and in Pb+Pb collisions \cite{Santo:1994um,Aggarwal:2000th} also
raised high hope of detecting QGP. However, it was later found that 
the   SPS energy data are well explained in models without QGP 
\cite{Chaudhuri:1995se,Chaudhuri:2000az}.

In the present calculation we have used the bag model equation of state for the QGP phase, 
$p_{q}=a_{q}T^{4}-B$ with $a_{q}=42.25\pi ^{2}/90$. The hadronic equation of state was generalized to include all the mesonic resonances with mass $<$ 2 GeV. The cut off 2 GeV is rather arbitrary and we verify that the results do not depend on the value of cut-off significantly. The bag constant $B$ was obtained from the Gibbs condition $p_{QGP}(T_{c})=p_{had}(T_{c})$. The critical temperature ($T_{c}$) and the freeze-out temperature ($T_{F}$) are assumed to be 180 MeV and 100 MeV respectively.

Photon emission rate from QGP and from  hot hadronic gas are
well known.
For the single photons from hadronic gas we include the following processes,

(a) $\pi\pi \rightarrow \rho \gamma$, (b) $\pi \rho \rightarrow \pi \gamma$,
(c) $\omega \rightarrow \pi \gamma$, (d) $\rho \rightarrow \pi \pi \gamma$
(e) $\pi \rho \rightarrow A_1 \rightarrow \pi \gamma$

\noindent rates for which are calculated in 
\cite{Nadeau:1992cn,Xiong:1992ui}. Photon emission rates
from QGP are calculated in
\cite{Kapusta:1991qp,Aurenche:1998nw,Aurenche:1999tq,Aurenche:2000gf}.

Hydrodynamic models require initial time ($\tau_i$) and initial energy density profile ($\varepsilon_i(r)$). $\tau_i$ is essentially the thermalisation
time beyond which hydrodynamic is applicable.
RHIC data on elliptic flow as well as hadron $p_T$ 
spectra require very short time scale of thermalisation $\tau_i$=0.6 fm \cite{Kolb:2003dz,Kolb:2001qz}. In the first scenario, when QGP is produced in the initial state, we use this value as the initial time. The PHENIX (preliminary) direct photon data are 
in the 0-20\% centrality range of collisions. In this centrality range,  the average number of binary collisions is $<N_{binary}> \approx 779$. We have assumed that 0-20\% centrality Au+Au collision corresponds to
central collisions of nuclei $A_{eff}$, such that the average 
number of binary collisions is reproduced. With 
$T_{AA}(b=0) \approx \frac{A^2}{\pi R^2}=\frac{A^2}{\pi (1.12A^{1/3})^2}$, we obtain $A_{eff} \approx 143$. 
For the initial energy density profile we use a Woods-Saxon form with radius $R_{eff}$=5.85 fm (corresponding to nucleus $A_{eff}$=143). For the diffuseness parameter we use $a=0.54$ fm. The central value of 
the energy density could be obtained by fitting the PHENIX preliminary data. However, presently we donot attempt exact fit to the data. Rather
we have calculated the thermal photon yield with three values of
central energy density, $\varepsilon(r=0)$=27.8,47.1 and 75.1 $GeV/fm^3$
corresponding to initial temperature $T_i$=0.35, 0.40 and 0.45 GeV respectively.

\begin{figure}[h]
\centerline{\psfig{figure=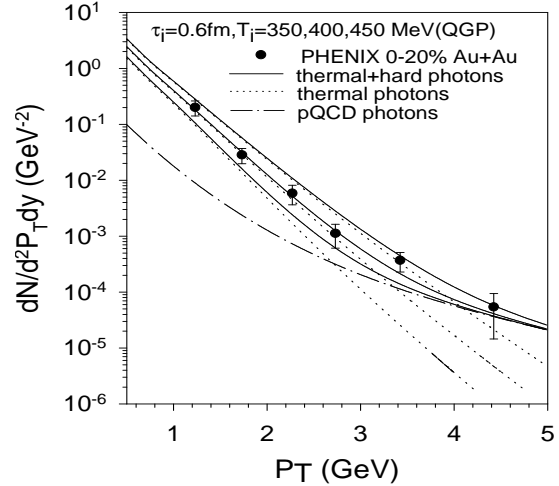,height=10cm,width=8cm}}
\vspace{-2.5cm}
\caption{PHENIX (preliminary) transverse momentum distribution of direct photons in 0-20\% centrality Au+Au collisions. The dotted lines are thermal
photons from QGP with initial temperature 350,400 and 450 MeV (bottom to top), thermalised at $\tau_i$=0.6 fm . The dash-dotted
line is the pQCD photons. Solid lines are the sum of the pQCD and thermal photons.}  
\end{figure}

Results of our calculation are shown in Fig.1. The three dotted lines
are the thermal photons from initial QGP at temperature 350,400 and
450 MeV respectively.  
In Fig.1, pQCD (hard) photons \cite{Jager:2002xm} are shown as the dash-dotted line.  Sum of thermal and pQCD photon yield are shown
as the solid lines.  pQCD photons alone can not explain the PHENIX  
data throughout  the $p_T$ range. It explains the data at large $p_T$ but underestimate the yield at low $p_T$. PHENIX (preliminary) direct photon data require thermal photons. Thermal photons dominate the photon spectra at
low $p_T$. As seen in Fig.1,
if the initial QGP is formed at temperature $T_i$=350 MeV, the PHENIX data are underpredicted. The data are over predicted for initial temperature 
$T_i$=450 MeV. PHENIX (preliminary) direct photon data are explained if the initial temperature is $T_i$=400 MeV. The analysis suggest that the PHENIX 
(preliminary) direct photon data in 0-20\% centrality
Au+Au collisions are explained if Au+Au collisions produce QGP with
central temperature 400 MeV at an initial time of 0.6 fm.

\begin{figure}[h]
\centerline{\psfig{figure=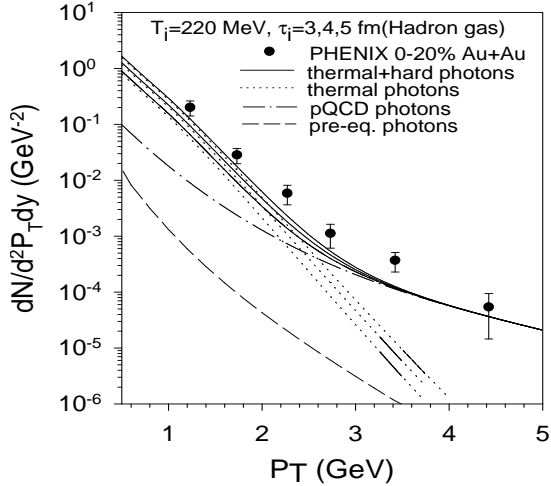,height=10cm,width=8cm}}
\vspace{-2.5cm}
\caption{PHENIX (preliminary) transverse momentum distribution of direct photons in 0-20\% centrality Au+Au collisions. The dotted lines are thermal
photons from initial hot hadronic gas at initial temperature of 220 MeV, thermalised at $\tau_i$=3,4 and 5 fm (from bottom to top). pQCD photons are shown as the dash-dotted
The solid lines are the sum of the pQCD and thermal photons. 
Photons from a pre-equilibrium QGP are shown as the dash-dot-dot line.}
\end{figure}

Let us now consider the photon yield in the second scenario when Au+Au collisions produce  hot hadronic gas. As it is
well known, photon emission rate from QGP and from hot hadronic gas
are very similar. Degeneracy of resonance hadron gas is also of the same order as that of a QGP. Thus it is expected that a hot hadronic gas thermalised at $\tau_i \sim$ 0.6 at an
initial temperature   of $T_i \sim$400 MeV will give a similar description to the PHENIX data, as it is obtained in the 1st scenario. However 
hadronic gas at such a high temperature is physically unacceptable. Density of the gas is very large $\rho_{hadron}  \sim 50 fm^{-3}$. 
Hadrons can not retain their identity at such a high density. What is the
acceptable limit of initial temperature of the hot hadronic gas? Recent
lattice QCD calculations indicate that critical temperature of 
confinement-deconfinement phase transition is $T_c=190 \pm 10$ MeV
\cite{Katz:2005br}. Possibly hadrons can retain their identity still at higher temperature. We choose $T_i$=220 MeV as the physically acceptable 
value for the
initial temperature of the hot hadronic gas. At this temperature density
of the hadron gas is $\rho_{hadrom} \sim 1.7 fm^{-3}$ just below the
limit  $\rho_{hadron} < 2 fm^{-3}$ so that the hadrons retain their identity.
 Now what should be the thermalisation
time scale for a hot hadronic gas? 
Will it be as small as that of a QGP? If the hadronic gas at initial
temperature of 220 MeV thermalises
at the same time scale (0.6 fm) the PHENIX (preliminary) direct photon data are not explained. However, thermalisation time scale for a hadronic gas can be 
longer.
Indeed, very small thermalisation time
scale ($\tau_i \sim 0.6 fm$) obtained for a QGP  is a puzzle in heavy ion physics and is not understood properly.
If the parton-parton collisions are responsible for thermalisation, the
thermalisation time is significantly longer. 
Calculations performed within the 'bottom-up' thermalisation scenario including $2 \leftrightarrow 2$ and
$2 \leftrightarrow 3$ processes estimate $\tau_i > 2.6 fm$ \cite{Baier:2000sb,Baier:2002bt}.
Only mechanism for short thermalisation scale is plasma instability in soft modes of the gluon field  \cite{Mrowczynski:2005ki}. Due to rapid longitudinal
expansion of the system, the  momentum spectrum of partons  quickly become anisotropic, the width of the longitudinal momentum distribution
become narrower that the width of the transverse transverse momentum distribution $<\Delta p^2_L> << <\Delta p^2_\perp>$.  A transverse chromo-electric field develop, which
further enhances the fluctuations in the parton momentum distribution. Anisotropic
parton momentum distribution can cause magnetic (transverse)
instabilities, which can be very efficient in thermalising the system.
Characteristic inverse time scale for instability development is roughly
of the order of $gT$ for sufficiently anisotropic momentum distribution. 
Thermalisation time scale of a hot hadronic gas is expected to be
greater than that of a QGP as it is likely that in a hot hadronic gas,
thermalisation will be driven by collisional process rather than by the
plasma instabilities. Plasma instabilities, in hot hadronic gas,
being electro-magnetic in nature, will not be efficient to thermalise the
system rapidly ($\alpha << \alpha_s$). 
In  \cite{Chaudhuri:2001bi} thermalisation of linear sigma model fields
were studied numerically. In the model, sigma model fields interact with
a heat bath. Irrespective of the initial field configuration, sigma model fields
thermalises in the time scale $\sim$ 5 fm.  In resonance hadronic gas 
thermalisation process will be faster as number
of fields greatly exceed that of the sigma model. A reasonable estimate will be $\tau_i \sim$ 1-5 fm. 

In Fig.2 photon yield from the hot hadronic gas, with initial temperature 
$T_i$=220 MeV for three time scale of thermalisation, $\tau_i$=3,4, and 5
fm are shown (the dotted lines).  Even with large thermalisation scale, $\tau_i$=5 fm, PHENIX data are not explained if Au+Au collisions
produces hot hadronic gas at an initial temperature of 220 MeV. The data
remain underpredicted by a factor of two or more.
The hadronic state can not produce the required number of photons.  

With large thermalisation time,
$\tau_i$ =5 fm, the fluid matter spends considerable time in the 
pre-equilibrium stage. Thus while pre-equilibrium photons may not be important if Au+Au collisions produces QGP ($\tau_i$=0.6 fm), they may
be important if the collisions lead to hot hadronic gas formation
($\tau_i$=5 fm). 
To unequivocally reject the hot hadronic gas scenario
it is important to estimate the pre-equilibrium photons.
Unfortunately emission of photons from a pre-equilibrium hadronic gas is not studied. However, photon emission 
from a pre-equilibrium QGP has been studied earlier
 \cite{Traxler:1995kx,Chaudhuri:1998kv}. Pre-equilibrium photons are order of magnitude less than
equilibrium photons.  Following \cite{Chaudhuri:1998kv} we have
estimated the photon yield from a  chemically non-equilibrated partonic system with initial conditions dictated by the
HIJING simulation for RHIC Au+Au collisions. The partonic system 
achieved kinetic equilibrium by the time $\tau_{iso}$=0.31 fm  at temperature of
570 MeV. At $\tau_{iso}$=0.31 fm, gluon and quarks fugacities are 0.09 and 0.02 respectively. We have considered only longitudinal expansion. 
In Fig.2, pre-equilibrium photons
(integrated over the time scale 0.31 to 5 fm) are shown as the dashed line.
Photons from pre-equilibrium stage contribute insignificantly. The reason can be understood
easily. The photon emission rate is weighted by the fugacities which
remain at low values in the time scale integrated. 
With transverse expansion, pre-equilibrium emission at
large $p_T$ will increase, however, the increase would never be
large enough to reckon them. 
If the pre-equilibrium hot hadronic gas contribute to the same order as
the pre-equilibrium QGP, we can safely ignore them. Even if pre-equilibrium
photons from the hot hadronic gas exceed that from the pre-equilibrium QGP  by a factor of 10, they still can be neglected.
   We conclude that PHENIX (preliminary) direct photon data
are not explained if Au+Au collisions produces a physically acceptable 
hot hadronic gas, with initial temperature less than 220 MeV at an
initial time less than 5 fm.

To summarise, we have analysed   the preliminary PHENIX data on the
transverse momentum distribution of direct photons in 0-20\% centrality 
Au+Au collisions at $\sqrt{s_{NN}}$=200 GeV. Two scenarios are considered, (i) Au+Au collisions produces a QGP and (ii) Au+Au collisions produces hot hadronic gas.  PHENIX (preliminary) direct photon data are explained in the first scenario if QGP is produced at initial time $\tau_i$=0.6 fm at an initial temperature  $T_i$=400 MeV. 
PHENIX data are not explained in the alternate scenario if the hot hadronic
gas is produced with initial temperature less or equal to 220 MeV at an
initial time 5 fm or less. The data remain underpredicted. As the
 hot hadronic
gas at higher temperature or with longer thermalisation time is physically
unacceptable, we  conclude that PHENIX (preliminary) direct photon data are  explained only if
QGP is produced in 0-20\% centrality Au+Au collisions.

\end{document}